\documentclass[12pt]{article}

\usepackage{amsmath}
\usepackage{amssymb}
\usepackage[dvips]{graphicx}

\usepackage{cite}

\makeatletter
 
  \@addtoreset{equation}{section}
 \makeatother

\newcommand {\n}{\nonumber \\}
\newcommand {\tr}{\mbox{tr}}

\begin{document}
\setlength{\oddsidemargin}{0cm}
\setlength{\baselineskip}{7mm}

\begin{titlepage}

~~\\

\vspace*{0cm}
    \begin{Large}
       \begin{center}
         {Three-Algebra BFSS Matrix Theory}
       \end{center}
    \end{Large}
\vspace{1cm}

\begin{center}
           Matsuo S{\sc ato}\footnote
           {
e-mail address : msato@cc.hirosaki-u.ac.jp}\\
      \vspace{1cm}
       
         {\it Department of Natural Science, Faculty of Education, Hirosaki University\\ 
 Bunkyo-cho 1, Hirosaki, Aomori 036-8560, Japan}

\end{center}

\hspace{5cm}

\begin{abstract}
\noindent

We extend the BFSS matrix theory by means of Lie 3-algebra. The extended model possesses the same supersymmetry as the original BFSS matrix theory, and thus as the infinite momentum frame limit of M-theory. We study dynamics of the model by choosing the minimal Lie 3-algebra that includes u(N) algebra. We can solve a constraint in the minimal model and obtain two phases. In one phase, the model reduces to the original matrix model. In another phase, it reduces to a simple supersymmetric model.

\end{abstract}

\vfill
\end{titlepage}
\vfil\eject

\setcounter{footnote}{0}

\section{Introduction}
\setcounter{equation}{0}
The BFSS matrix theory \cite{BFSS} is a promising matrix model for the infinite momentum frame (IMF) limit of M-theory. This model owns the same space-time supersymmetry as the IMF limit of M-theory. The existence of graviton is demanded by the thirty-two supercharges that generate the $\mathcal{N}=1$ supersymmetry in eleven dimensions.

In spite of the fact that this matrix model is known to describe some dynamics in M-theory, it is hard to analyze the model owing to its many interactions. Many ideas are necessary to study M-theory. Extending the matrix model is one reasonable way to obtain new ideas to study not only M-theory but the original BFSS matrix theory.

Recently, 3-algebraic symmetries were found in multiple M2-brane effective actions \cite{BLG1, Gustavsson, BLG2, Lorentz0, Iso, ABJM, N=6BL}
\footnote{ABJM theory can also be rewritten as a 3-algebra manifest form \cite{N=6BL}.} and 3-algebras have been intensively studied \cite{Filippov, BergshoeffSezginTownsend, deWHN, Nogo1, Jabbari, GomisSalimPasserini, HosomichiLeeLee, Nogo2, sp1, sp2, Nogo3, Lorentz1, Lorentz2, Lorentz3, sp3, PangWang, ABJ, text, Nogo8, NishinoRajpoot, kac, Class, GustavssonRey, HanadaMannelliMatsuo, IshikiShimasakiTsuchiya, KawaiShimasakiTsuchiya, IshikiShimasakiTsuchiya2, DeBellisSaemannSzabo, Palmkvist}. Originally, the bosonic part of the membrane action has a 3-algebraic symmetry. That is, it can be written in the symmetry manifest form as $S=T_{M2} \int d^3 \sigma \sqrt{g}\left(-\frac{1}{12}(\frac{1}{\sqrt{g}}\{ X^L, X^M, X^N \})^2+\Lambda\right)$ where $\{\,, \,, \, \}$ denotes Nambu-Poisson bracket \cite{Nambu, Yoneya}. Therefore, one can expect that 3-algebraic symmetry plays important roles in M-theory\footnote
{
A formulation of M-theory by a cubic matrix action was proposed by Smolin \cite{Smolin1, Smolin2, Azuma} } \cite{Minic, bosonicM, MModel, LorentzianM, ZariskiM}. 

In this paper, we make an extension of the BFSS matrix theory. "Extention" implies that the model owns a 3-algebraic structure that includes a Lie-algebraic structure. The model allows any 3-algebra whose triple product is totally antisymmetric. Such 3-algebra is so-called Lie 3-algebra. It owns the same supersymmetry as the BFSS matrix theory, that is, the same supersymmetry as the IMF limit of M-theory.  Therefore, gravitons exist in the extended model. 

We also investigate dynamics of the model by choosing the minimal Lie 3-algebra that includes $u(N)$ algebra. In the minimal model, we can solve a constraint and obtain two phases. In one phase, the model reduces to the original BFSS matrix theory. In another phase, the model reduces to a simple supersymmetric model, which is easier to analyze.

The rest of this paper is organized as follows. In section 2, we review on Lie 3-algebra. In section 3, we construct an extended BFSS matrix theory that allows any Lie 3-algebra. The algebra of the supersymmetry closes on-shell. In section 4, we study the model with the minimal Lie 3-algebra that includes $u(N)$. In section 5, we study a phase structure of the model. In section 6, we make a conclusion and discusssions.

\vspace{1cm}

\section{Lie 3-algebra}
Lorentzian Lie 3-algebra includes Lie-algebra. In this section, we review on this algebra. 

Lie 3-algebra is expressed by three-brackets,
\begin{equation}
[T^A, T^B, T^C]= f^{A B C}_{\qquad D} T^D,
\end{equation}
where $f^{A B C}_{\qquad D} $ are structure constants. The bracket is totally antisymmetric. The gauge transformation is given by 
\begin{equation}
\delta X = \Lambda_{AB} [T^A, T^B, X].
\end{equation}
An inverse of a metric is given by symmetric products,
\begin{equation}
g^{AB}=<T^AT^B>
\end{equation}
and indices $A, B, \cdots$ are raised and lowered by $g^{AB}$ and $g_{AB}$.
For gauge covariance, the algebra must satisfy the gauge invariance of the metric and the fundamental identity. Gauge invariance of the metric implies that
\begin{equation}
f^{ABCD}=-f^{ABDC}.
\end{equation}
Thus, indices of $f^{ABCD}$ are totally anti-symmetric.
The fundamental identity, which is an analogue of Jacobi identity is given by
\begin{equation}
\delta [X, Y, Z]=[\delta X, Y, Z]+[X, \delta Y, Z]+[X, Y, \delta Z].
\end{equation} 
This is expressed in terms of the structure constants as
\begin{equation}
f^{CDE}_{\qquad F}f^{ABFG}
=f^{ABC}_{\qquad  F}f^{FDEG}
+f^{ABD}_{\qquad  F}f^{CFEG}
+f^{ABE}_{\qquad  F}f^{CDFG}.
\end{equation}

In addition, Lorentzian Lie 3-algebra owns a maximally isotropic center: Generators are given by $T^{\alpha}$, $T^{\bar{\alpha}}$ and $T^{I}$ where $T^{\bar{\alpha}}$ are elements of the center. Non-zero components of a metric are given by $g_{\alpha \bar{\beta}}=-\delta_{\alpha \beta}$ and positive definite metric $g_{IJ}$. 

Finite-dimensional indecomposable metric Lie 3-algebras with maximally isotropic center are categorised in \cite{Class}. 
The vector space is defined by
\begin{equation}
V= \bigoplus_{\alpha=1}^{R}(\mathbb{R} T^{\alpha} \oplus \mathbb{R} T^{\bar{\alpha}}) 
\oplus
\bigoplus_{s=1}^{K} W_s
\oplus
\bigoplus_{\pi=1}^{L} E_{\pi}
\oplus
E_0,
\end{equation}
where each $W_s$ stands for a vector space of a Lie algebra. $E_{\pi}$ and $E_0$ are positive definite vector spaces with dim $E_{\pi}$ =2 and dim $E_0 \leqq {R \choose 3}$. The elements $T^{\alpha}$, $T^{\bar{\alpha}}$, $T^{i_s} (\in W_s)$, $T^{a_{\pi}} (\in E_{\pi})$ and $T^p (\in E_0)$ satisfy the following algebra,
\begin{eqnarray}
&&[T^{\alpha}, T^{i_s}, T^{j_s}]= \kappa^{\alpha}_{s} f_{s \quad k_s}^{i_s j_s}T^{k_s} \n
&&[T^{i_s}, T^{j_s}, T^{k_s}]= -f_s^{i_s j_s k_s} \kappa_{s \bar{\alpha}}T^{\bar{\alpha}} \n
&&[T^{a_{\pi}}, T^{\alpha}, T^{\beta}]= J_{\pi}^{\alpha \beta} \epsilon^{a_{\pi}}_{\,\,\,\,\,\,\, b_{\pi}}T^{b_{\pi}} \n
&&[T^{\alpha}, T^{a_{\pi}}, T^{b_{\pi}}]= J_{\pi \, \bar{\beta}}^{\alpha} \epsilon^{a_{\pi} b_{\pi}} T^{\bar{\beta}} \n
&&[T^{\alpha}, T^{\beta}, T^{\gamma}]= K^{\alpha \beta \gamma}_{\quad \,\,\, p}T^{p} + L^{\alpha \beta \gamma}_{\quad \,\,\, \bar{\delta}} T^{\bar{\delta}} \n
&&[T^{p}, T^{\alpha}, T^{\beta}]= K^{p \alpha \beta}_{\quad \,\,\, \bar{\gamma}}T^{\bar{\gamma}},
\label{full3-alg}
\end{eqnarray}  
where $J_{\pi}^{\alpha \beta} = 
\eta_{\pi}^{\alpha} \zeta_{\pi}^{\beta}
-
\eta_{\pi}^{\beta} \zeta_{\pi}^{\alpha}$.
$\eta_{\pi}^{\alpha}$, $\zeta_{\pi}^{\alpha}$ and $\kappa_s^{\alpha}$ are arbitrary constant vectors and $L^{\alpha \beta \gamma \delta}$ is an arbitrary constant totally antisymmetric tensor. $K^{\alpha \beta \gamma}_{\quad \,\,\, p}$ is totally antisymmetric in $\alpha$, $\beta$ and $\gamma$ and satisfies
\begin{equation}
K^{\alpha \beta \gamma p} K^{\delta \epsilon \phi}_{\quad \,\,\, p}
-
K^{\alpha \beta \delta p} K^{\epsilon \phi \gamma}_{\quad \,\,\, p}
+
K^{\alpha \beta \epsilon p} K^{\phi \gamma \delta}_{\quad \,\,\, p}
-
K^{\alpha \beta \phi p} K^{\gamma \delta \epsilon}_{\quad \,\,\, p}
=0.
\end{equation}
$f_{s \quad k_s}^{i_s j_s}$ are structure constants of simple Lie-algebras and $\epsilon^{a_{\pi} b_{\pi}}$ are antisymmetric tensors in two-dimensional vector spaces. Non-zero metrics are given by $g_{\alpha \bar{\beta}}=-\delta_{\alpha \beta}$ and positive definite metrics $g_{i_s j_s}$, $g_{a_{\pi} b_{\pi}}$ and $g_{pq}$. 

An infinite-dimensional Lorentzian Lie 3-algebra associated with Kac-Moody algebra was studied in \cite{kac}.

\section{Extended BFSS Matrix Theory}
In this section, we extend the BFSS matrix theory to a Lie 3-algebraic model. We consider a scalar $\Phi$, SO(9) vector $X^a$ $(a = 1, \cdots 9)$ and SO(1,9) Majorana-Weyl fermion $\Theta$ spanned by Lie 3-algebra.
The Weyl condition is given by 
\begin{equation}
\Gamma^{10}\Theta=-\Theta.
\end{equation}A covariant derivative in a Lie 3-algebraic gauge theory in one dimension is defined as \cite{BLG1, Gustavsson, BLG2}
\begin{equation}
D_0 X= \partial_0 X -ia_{0AB}[T^A, T^B, X],
\end{equation}
where $a_{0AB}$ is a gauge field in one dimension. The gauge transformation is given by 
\begin{eqnarray}
&&\delta X =\Lambda_{\beta \gamma} [T^{\beta}, T^{\gamma}, X] \n
&&\delta A_{0 \,\, A}^B= -i\bigl(\partial_0 \tilde{\Lambda}^B_{\,\,\,\,\, A}
-i(A_{0 \,\, A}^C \tilde{\Lambda}^B_{\,\,\,\,\, C} -\tilde{\Lambda}^C_{\,\,\,\,\, A}A_{0 \,\, C}^B) \bigr)
\end{eqnarray}
where $A_{0 \,\, A}^B=a_{0 \, CD}f^{CDB}_{\qquad A}$ and $\tilde{\Lambda}^B_{\,\,\,\,\, A}= \Lambda_{CD} f^{CDB}_{\qquad A}$.

We extend the dynamical supertransformation of the BFSS matrix theory as follows,
\begin{eqnarray}
&& \delta a^0 = \frac{i}{2} (\Phi \bar{E} \Gamma^0 \Theta -\bar{E} \Gamma^0 \Theta \Phi) \n 
&& \delta X^a = i \bar{E} \Gamma^a \Theta \n
&& \delta \Phi =0 \n
&& \delta \Theta = (-D_0 X_a \Gamma^{0a} + \frac{i}{2} [\Phi, X_a, X_b ] \Gamma^{ab}) ,
\label{SUSY2}
\end{eqnarray}
where $a$, $b$ run from 1 to 9.
We consider a theory in which $\Phi$ is covariantly constant,
\begin{equation}
D_0 \Phi = 0. 
\label{constraint}
\end{equation}
This condition is necessary for closure of the supersymmetry algebra. This condition is consistent with the supertransformation, namely 
\begin{equation}
\delta D_0 \Phi = 0.
\end{equation}
From this transformation, we obtain a supersymmetry algebra,
\begin{eqnarray}
&& (\delta_2 \delta_1 - \delta_1 \delta_2) \Phi =0 \n
&& (\delta_2 \delta_1 - \delta_1 \delta_2) X^a 
=-2i \bar{E}_2 \Gamma^0 E_1 D_0 X^a
+2 \bar{E}_2 \Gamma^b E_1 [ \Phi, X_b, X^a] \n
&& (\delta_2 \delta_1 - \delta_1 \delta_2) A_{0 \,\, A}^B= -i\bigl(\partial_0 \tilde{\Lambda}^B_{\,\,\,\,\, A}
-i(A_{0 \,\, A}^C \tilde{\Lambda}^B_{\,\,\,\,\, C} -\tilde{\Lambda}^C_{\,\,\,\,\, A}A_{0 \,\, C}^B) \bigr) \n
&& (\delta_2 \delta_1 - \delta_1 \delta_2) \Theta
= -2i \bar{E}_2 \Gamma^0 E_1 D_0 \Theta
+2 \bar{E}_2 \Gamma^b E_1 [\Phi, X_b, \Theta] \n
&& \qquad \qquad \qquad \quad +(\frac{7}{8} \bar{E}_2 \Gamma_{\mu} E_1 \Gamma^{\mu} 
- \frac{1}{16} \bar{E}_2 \Gamma_{\mu_1 \cdots \mu_5} E_1 \Gamma^{\mu_1 \cdots \mu_5})
(i \Gamma^0 D_0 \Theta + [\Phi, \Gamma^b \Theta_b, \Theta]), 
\end{eqnarray}
where $\Lambda_{AB}=\bar{E}_2 \Gamma^a E_1 (\Phi_A X_{aB} -X_{aA} \Phi_B)$.
This algebra closes among the supertransformation, translation and gauge transformation on-shell if we take 
\begin{equation}
i \Gamma^0 D_0 \Theta + [\Phi, \Gamma^b \Theta_b, \Theta]=0
\label{eomTheta}
\end{equation}
as an equation of motion of the fermion. By transforming this, we obtain a part of equations of motion of bosons.
\begin{eqnarray}
&& D_0^2 X^a + [\Phi, X_b, [\Phi, X^b, X^a]] 
-\frac{1}{2}[ \Phi, \bar{\Theta} \Gamma^0, \Theta] =0 \label{eomXa} \\
&& i[\Phi, X_a, D_0 X^a] -\frac{1}{2} [ \Phi, \bar{\Theta} \Gamma^0, \Theta]=0
\label{eomA}
\end{eqnarray}
One can show that an action
\begin{equation}
S=\int d\tau < \frac{1}{2} (D_0 X^a)^2 + \frac{1}{4}[\Phi, X_a, X_b]^2 
-\frac{i}{2} \bar{\Theta} \Gamma^0 D_0 \Theta -\frac{1}{2} \bar{\Theta}[ \Phi, \Gamma^a X_a, \Theta]> \label{preBFSSext}
\end{equation}  
with a constraint (\ref{constraint}) is invariant under (\ref{SUSY2}). One can also reproduce the equations of motion of $\Theta$ (\ref{eomTheta}) and $X^a$ (\ref{eomXa}). The equation of motion of the gauge field is given by
\begin{equation}
i[O, D_0 X^a, X_a] + \frac{1}{2}[O, \bar{\Theta}, \Gamma^0 \Theta]=0, 
\end{equation}
where $O$ is an arbitrary field. If we take $\Phi$ as $O$, (\ref{eomA}) is reproduced. 

Because the kinetic term $-\partial_0 X^a_{\bar{\alpha}} \partial_0 X_{a \alpha}$ has wrong sign, there are ghosts in this action. Thus, we introduce an additional shift symmetry and gauge away $X^a_{\bar{\alpha}}$ and $\Theta_{\bar{\alpha}}$. Such a prescription is given in \cite{Ghost-Free}. The action (\ref{preBFSSext}) has a global shift symmetry of the center field,
\begin{eqnarray}
&& \delta x_{\bar{\alpha}}^a=\Lambda^a_{\bar{\alpha}} \n
&& \delta \Theta_{\bar{\alpha}} = \xi_{\bar{\alpha}}.
\end{eqnarray}
In order to make it to a local symmetry, we introduce new fields, SO(9) vectors
$C^a_{\bar{\alpha}}$ and SO(1,9) Majorana Weyl fermions $\chi_{\bar{\alpha}}$. The unitary action is given by\begin{eqnarray}
S&=&\int d\tau \biggl(< \frac{1}{2} (D_0 X^a)^2 + \frac{1}{4}[\Phi, X_a, X_b]^2 
-\frac{i}{2} \bar{\Theta} \Gamma^0 D_0 \Theta -\frac{1}{2} \bar{\Theta}[ \Phi, \Gamma^a X_a, \Theta]> \n
&& \qquad +\bar{\theta}_{\alpha} \chi_{\bar{\alpha}} + (\partial_0 X^a_{\alpha})C^a_{\bar{\alpha}} \biggr) 
\label{unitaryaction}
\end{eqnarray} 
with (\ref{constraint}).
This action is invariant under the additional local shift symmetry,
\begin{eqnarray}
&& \delta x_{\bar{\alpha}}^a=\Lambda^a_{\bar{\alpha}} \n
&& \delta \Theta_{\bar{\alpha}} = \xi_{\bar{\alpha}} \n
&& \delta C^a_{\bar{\alpha}} = \partial_0 \Lambda^a_{\bar{\alpha}} \n
&& \delta \chi_{\bar{\alpha}} = -i \Gamma^0 \partial_0 \xi_{\bar{\alpha}}.
\label{newgauge}
\end{eqnarray}

The dynamical supertransformation under which (\ref{unitaryaction}) with (\ref{constraint}) is invariant is given by (\ref{SUSY2}) and 
\begin{eqnarray}
&& \delta C^a_{\bar{\alpha}} = \bar{E} \Gamma^a \Gamma_0 \chi_{\bar{\alpha}} \n
&& \delta \chi_{\bar{\alpha}} = -i \partial_0 C^a_{\bar{\alpha}} \Gamma^a E.
\end{eqnarray}
From this transformation, we obtain 
\begin{eqnarray}
&& (\delta_2 \delta_1 - \delta_1 \delta_2) \Phi =0 \n
&& (\delta_2 \delta_1 - \delta_1 \delta_2) X^a 
=-2i \bar{E}_2 \Gamma^0 E_1 D_0 X^a
+2 \bar{E}_2 \Gamma^b E_1 [ \Phi, X_b, X^a] \n
&& (\delta_2 \delta_1 - \delta_1 \delta_2) a_0
= -iD_0(2 \bar{E}_2 \Gamma^a E_1 \frac{1}{2}(\Phi X_a- X_a \Phi)) \n
&& (\delta_2 \delta_1 - \delta_1 \delta_2) \Theta
= -2i \bar{E}_2 \Gamma^0 E_1 D_0 \Theta
+2 \bar{E}_2 \Gamma^b E_1 [\Phi, X_b, \Theta] \n
&& \qquad \qquad \qquad \quad +(\frac{7}{8} \bar{E}_2 \Gamma_{\mu} E_1 \Gamma^{\mu} 
- \frac{1}{16} \bar{E}_2 \Gamma_{\mu_1 \cdots \mu_5} E_1 \Gamma^{\mu_1 \cdots \mu_5})
(i \Gamma^0 D_0 \Theta + [\Phi, \Gamma^b \Theta_b, \Theta] + \chi_{\bar{\alpha}} T^{\bar{\alpha}}) \n
&& \qquad \qquad \qquad \quad -(\frac{7}{8} \bar{E}_2 \Gamma_{\mu} E_1 \Gamma^{\mu} 
- \frac{1}{16} \bar{E}_2 \Gamma_{\mu_1 \cdots \mu_5} E_1 \Gamma^{\mu_1 \cdots \mu_5}) \chi_{\bar{\alpha}} T^{\bar{\alpha}} \n
&& (\delta_2 \delta_1 - \delta_1 \delta_2) C_{\bar{\alpha}}^a 
= \partial_0 (-2i \bar{E}_1 \Gamma^0 E_2 C_{\bar{\alpha}}^a) \n
&& (\delta_2 \delta_1 - \delta_1 \delta_2) \chi_{\bar{\alpha}}
= -i \Gamma^0 \partial_0 (16 \bar{E}_2 \Gamma_0 E_1 \Gamma^0 \chi_{\bar{\alpha}}).
\label{BFSSalg1}
\end{eqnarray}
The algebra closes among the supertransformation, the translation and the gauge transformations on-shell.

This model also has kinematical supersymmetry,
\begin{equation}
\tilde{\delta} \Theta = \tilde{E}
\end{equation}
The algebra is given by
\begin{eqnarray}
&&(\tilde{\delta}_2 \tilde{\delta}_1- \tilde{\delta}_1 \tilde{\delta}_2) \Phi = 0 \n
&&(\tilde{\delta}_2 \tilde{\delta}_1- \tilde{\delta}_1 \tilde{\delta}_2) X^M =0 \n
&&(\tilde{\delta}_2 \tilde{\delta}_1- \tilde{\delta}_1 \tilde{\delta}_2) \Theta =0
\label{BFSSalg2}
\end{eqnarray}
and
\begin{eqnarray}
&&(\tilde{\delta}_2 \delta_1- \delta_1 \tilde{\delta}_2) \Phi = 0 \n
&&(\tilde{\delta}_2 \delta_1- \delta_1 \tilde{\delta}_2) X^M = i \bar{E}_1 \Gamma^M E_2 \n
&&(\tilde{\delta}_2 \delta_1- \delta_1 \tilde{\delta}_2) \Theta = 0.
\label{BFSSalg3}
\end{eqnarray}

This supersymmetry algebra (\ref{BFSSalg1}), (\ref{BFSSalg2}) and (\ref{BFSSalg3}) is consistent with that of M-theory as in the same way as the BFSS matrix model \cite{MSUSY}:
\begin{eqnarray}
&&\{\tilde{Q}, Q^T\}=-2P^a \gamma_a \n
&&\{\tilde{Q}, \tilde{Q}^T\}=-2\sqrt{2}P^+ \n
&&\{Q, Q^T\}=2\sqrt{2}P^-
\end{eqnarray}

%
%
%

By using the additional shift symmetry (\ref{newgauge}), we can fix the gauge,
\begin{eqnarray}
&& X^a_{\bar{\alpha}}=0 \n
&& \Theta_{\bar{\alpha}}=0.
\label{ghostfree}
\end{eqnarray}
As a result, we obtain a ghost free theory (\ref{unitaryaction}) with (\ref{constraint}) and (\ref{ghostfree}).

\section{Minimal Model}

In this section, we study the unitary model in the previous section with the minimal Lorentzian Lie 3-algebra that includes $u(N)$ algebra. Non-zero structure constants are given by
\begin{equation}
f^{\alpha ijk}= \kappa^{\alpha} f^{ijk},
\end{equation}
where $\kappa^{\alpha}$ is an arbitrary vector and $f^{ij}_{\,\,\,\,\, k}$ are structure constants of $u(N)$ algebra. The other $f^{ABCD}$  are zero except for the antisymmetrized above form. Non-zero components of the metric are given by 
\begin{eqnarray}
&&g_{\alpha \bar{\beta}} = -\delta_{\alpha \beta} \n
&&g_{ij} = h_{ij},
\end{eqnarray}
where $h_{ij}$ is Cartan metric of the Lie algebra. Then, non-zero commutators are given by
\begin{eqnarray}
&& [T^{\alpha}, T^i, T^j]=\kappa^{\alpha}[T^i, T^j] \n
&& [T^i, T^j, T^k]=-f^{ijk}\kappa_{\bar{\alpha}}T^{\bar{\alpha}}.
\end{eqnarray}
As one can see from the above algebra, the Lorentzian direction is essentially one-dimensional. Thus in the following, we only consider the case $\kappa^{\alpha}$ is a one-dimensional vector. We denote $\alpha$ as $0$ and $\bar{\alpha}$ as $-$ and choose $\kappa^{\alpha}=1$. To summarize, the non-zero algebra is given by 
 \begin{eqnarray}
&&[T^0, T^i, T^j]=[T^i, T^j]=f^{ij}_{\quad k} T^k \n
&&[T^i, T^j, T^k] = f^{ijk} T^{-},
\end{eqnarray}
where $[T^i, T^j]$ is the Lie bracket. Non-zero components of the metric are given by
\begin{equation}
g_{-0}=-1, \quad g_{ij}=h_{ij}.
\end{equation}

 The covariant derivative in the Lie-algebra manifest form is given by 
\begin{equation}
D_0 X=\partial_0 X -i[\tilde{a}_0, X] -ia^{'i}X_iT^{-},
\end{equation}
where $\tilde{a}_0 \equiv 2a_{00i}T^i$ and $a^{'i}=a_{0jk}f^{jki}$.
The explicit form of the action (\ref{unitaryaction}) with (\ref{ghostfree}) is given by
\begin{eqnarray}
S&=& \int d\tau \Biggl( \tr \biggl( 
\frac{1}{2}(\tilde{D}_0 X^a)^2
+\frac{1}{4} (\Phi_0)^2 [X^{a}, X^{b}]^2 
+\frac{1}{2} (X_0^{a})^2[\Phi, X^{b}]^2 
- \frac{1}{2}(X_0^{a}[X_{a}, \Phi])^2 \n
&& \qquad \quad + \Phi_0 X_0^{a} [X_{a}, X_{b}][X^{b}, \Phi] \n
&& \qquad \quad -\frac{i}{2}\bar{\Theta} \Gamma^0 \tilde{D}_0 \Theta
-\frac{1}{2} \Phi_0 \bar{\Theta} \Gamma_{a}[X^{a}, \Theta]
+\frac{1}{2} X_0^{a} \bar{\Theta} \Gamma_{a}[\Phi, \Theta]
- \frac{1}{2} \bar{\Theta} \Gamma_{a} \Theta_0 [\Phi, X^{a}] 
+\frac{1}{2} \bar{\Theta}_0 \Gamma_{a} \Phi [X^{a}, \Theta]
\biggr) \n
&& \qquad \quad + i \partial_0 X^a_0 a_0^{'i} X_i^a 
- i (\tilde{D}_0 X^a)_i X^a_0 a_0^{'i} 
-\frac{1}{2} (X_o^a)^2 (a^{`i}_{0})^2
+\frac{1}{2}\bar{\Theta}_0 \Gamma^0 a_0^{'i} \Theta_i
-\frac{1}{2} \bar{\Theta}_i \Gamma^0 a_o^{'i} \Theta_0 \n
&& \qquad \quad + \bar{\Theta}_0 \chi 
+ \partial_0 X^a_0 C_{0a} 
\Biggr),
\end{eqnarray}
where $u(N)$ covariant derivative is defined as
\begin{equation}
\tilde{D}_0 X^a = \partial X^a -i [\tilde{a}_0, x^a].
\end{equation}
$\tilde{D}_0 \Theta$ and $\tilde{D}_0 \Phi$ are defined as in the same way.

Because $T^0$ component of the constraint (\ref{constraint}) is given by
\begin{equation}
\partial_0 \Phi_0=0,
\end{equation}
we can treat $\Phi_0$ as an order parameter. 
In this section, we consider $\Phi_0 \neq 0$ phase. 

Let us consider gauge symmetry of the action. The 3-algebra manifest form is given by 
\begin{equation}
\delta X_{\alpha} = \Lambda_{\beta \gamma} f^{\beta \gamma \delta}_{\quad \alpha} X_{\delta},
\end{equation}
where $X$ stands for any field, such as $X^M$, $\Phi$ and $\Theta$. The Lie algebra manifest form is given by
\begin{eqnarray}
&& \delta X_0 =0 \n
&& \delta X_i =  \Lambda^{(1) k}_{\quad \,\,\, i} X_k +\Lambda_i^{(2)} X_0, 
\end{eqnarray}
where 
$\Lambda^{(1) k}_{\quad \,\,\, i}=2 \Lambda_{0j} f^{jk}_{\quad i}$ and 
$\Lambda_i^{(2)}=\Lambda_{jk} f^{jk}_{\quad i}$ are independent gauge parameters. $\Lambda^{(1)}$ parametrizes the u(N) gauge transformation, whereas $\Lambda^{(2)}$ parametrizes a shift transformation. In $\Phi_0 \neq 0$ phase, by using the shift symmetry,  we can fix the gauge
\begin{eqnarray}
\Phi_i =0.
\end{eqnarray}
In this gauge, $D_0 \Phi =0$ can be solved as follows
\begin{eqnarray}
&& \partial_0 \Phi_0 = 0 \n
&& \partial_0 \Phi_{-}=0 \n
&& a_{0 i j}[T^i, T^j, T^0]=0
\end{eqnarray}
As a result, 
\begin{equation}
D_0 X^a=\partial_0 X^a_0 T^0 + \tilde{D}_0 X^a,
\end{equation}
Then, we obtain
\begin{equation}
S= \int d\tau \Bigl( \tr \bigl(
\frac{1}{2}(\tilde{D}_0 X^a)^2) 
+\frac{1}{4} (\Phi_0)^2 [X_a, X_b]^2
-\frac{i}{2} \bar{\Theta} \Gamma^0 \tilde{D}_0 \Theta
-\frac{1}{2} \Phi_0 \bar{\Theta} \Gamma^a [X_a, \Theta]
\bigr)
+ \bar{\Theta}_0 \chi_{-} + \partial_0 X^a_0 C^a_{-})
\Bigr)
\end{equation}
Because $\chi_{-}$, $C^a_{-}$, $\Theta_0$ and $X^a_0$ appear only in the last two terms, $\chi_{-}$ and $C^a_{-}$ can be integrated out and the last two terms vanishes. Therefore, the action reduces to the BFSS matrix theory
\begin{equation}
S= \int d\tau \tr \bigl(
\frac{1}{2}(\tilde{D}_0 X^a)^2
+\frac{1}{4} [X_a, X_b]^2
-\frac{i}{2} \bar{\Theta} \Gamma^0 \tilde{D}_0 \Theta
-\frac{1}{2} \bar{\Theta} \Gamma^a [X_a, \Theta]
\bigr)
\end{equation}
by appropriate field redefinitions and a scale transformation.

\section{New Phase}
In this section, we consider $\Phi_0 = 0$ phase. 
If we integrate out $\chi_{-}$ and $C^a_{-}$, we obtain conditions
\begin{eqnarray}
\Theta_0 =0 \n
\partial_0 x^a_0 =0.
\end{eqnarray}
Without loss of generality, we can choose
\begin{equation}
x^a_0 =v \delta^{a,9},
\end{equation}
where $v$ is a constant. Then, the action is rewritten as
\begin{eqnarray}
S= \int d\tau 
\Biggl(
\tr 
\biggl( 
&&\frac{1}{2}(\tilde{D}_0 X^a)^2
+\frac{1}{4} (\Phi_0)^2 [X^a, X^b]^2
+\frac{1}{2} v^2 [\Phi, X^m]^2
+\Phi_0 v [X^9, X^m][X_m, \Phi] \n
&& -\frac{i}{2} \bar{\Theta} \Gamma^0 \tilde{D}_0 \Theta
-\frac{1}{2} \Phi_0 \bar{\Theta} \Gamma_a [X^a, \Theta]
+ \frac{1}{2} v \bar{\Theta} \Gamma^9 [\Phi, \Theta]
\biggr) \n
&&-iv(\tilde{D}_0 X^9)_i a_0^{'i}
-\frac{1}{2} v^2 (a_{0 i}')^2
\Biggr),
\end{eqnarray}
where $m$ runs from 1 to 8.
In $\Phi_0 =0$ phase, we obtain
\begin{eqnarray}
S= \int d\tau 
\Biggl(
&&\tr 
\biggl(
\frac{1}{2}(\tilde{D}_0 X^a)^2
+\frac{1}{2} v^2 [\Phi, X^m]^2 
-\frac{i}{2} \bar{\Theta} \Gamma^0 \tilde{D}_0 \Theta
+ \frac{1}{2} v \bar{\Theta} \Gamma^9 [\Phi, \Theta]
\biggr) \n
&&-iv(\tilde{D}_0 X^9)_i a_0^{'i}
-\frac{1}{2} v^2 (a_{0 i}')^2
\Biggr),
\end{eqnarray}
and constraints become
\begin{eqnarray}
&&\partial_0 \Phi_{-}-i a^{'i}\Phi_i=0 \n
&&(\tilde{D}_0 \Phi)_i=0.
\end{eqnarray}
The first condition determines $\Phi_{-}$. Because the action does not depend on $\Phi_{-}$, only the second condition constrains the action.
In $v=0$ case, we have
\begin{equation}
S= \int d\tau 
\tr 
\biggl(
\frac{1}{2}(\tilde{D}_0 X^a)^2
-\frac{i}{2} \bar{\Theta} \Gamma^0 \tilde{D}_0 \Theta
\biggr) 
\end{equation}
There is no constraint. Because we can take $\tilde{a}_0=0$ gauge, the theory is free in this phase.
In $v \neq 0$ case, because the coefficient of $a_0^{'i}$ is constant and there is no constraint on $a_0^{'i}$, we can integrate out it. After integration, we obtain
\begin{equation}
S= \int d\tau 
\tr 
\biggl(
\frac{1}{2}(\tilde{D}_0 X^m)^2
+\frac{1}{2} v^2 [\Phi, X^m]^2 
-\frac{i}{2} \bar{\Theta} \Gamma^0 \tilde{D}_0 \Theta
+ \frac{1}{2} v \bar{\Theta} \Gamma^9 [\Phi, \Theta]
\biggr), \label{BFSSlastaction}
\end{equation}
with
\begin{equation}
\tilde{D}_0 \Phi=0.
\end{equation}
We should note that the action does not depend on $X^9$. $v$ can be absorbed by a redefinition of $\Phi$.

The dynamical supertransformation reduces to
\begin{eqnarray}
&& \delta \tilde{a}^0 =0 \n
&& \delta \Phi =0 \n
&& \delta X^m = i \bar{E} \Gamma^m \Theta \n
&& \delta \Theta = (- \tilde{D}_0 X_m \Gamma^{0m} 
+ i v [\Phi, X_m] \Gamma^{m9})E,
\end{eqnarray}
under which the action (\ref{BFSSlastaction}) is invariant. 

Supersymmetry algebra that follows from the above transformation is given by
\begin{eqnarray}
&&(\delta_2 \delta_1-\delta_1 \delta_2) \tilde{a}^0=0 \n
&&(\delta_2 \delta_1-\delta_1 \delta_2) \Phi =0 \n
&&(\delta_2 \delta_1-\delta_1 \delta_2) X^m 
= 2i \bar{E}_1 \Gamma^0 E_2  \tilde{D}_0 X^m
-2 v \bar{E}_1 \Gamma^9 E_2 [\Phi, X^m] \n
&& (\delta_2 \delta_1-\delta_1 \delta_2) \Theta
=-2i \bar{E}_2 \Gamma^0 E_1 \tilde{D}_0 \Theta
+2 \bar{E_2} \Gamma^9 E_1 v[\Phi, \Theta] \n
&& \qquad \qquad \qquad \qquad
+(i\bar{E}_2 \Gamma_{M} E_1 \Gamma^M 
- \frac{i}{4} \bar{E}_2 \Gamma_m E_1 \Gamma^m
+i \bar{E}_2 \Gamma_{M_1 \cdots M_5} E_1 \Gamma^{M_1 \cdots M_5} \n
&& \qquad \qquad \qquad \qquad \quad
-\frac{5}{4}i \bar{E}_2 \Gamma_{m M_1 M_2 M_3 M_4} E_1 \Gamma^{m M_1 M_2 M_3 M_4}
+ 10 i \bar{E}_2 \Gamma_{0 9 m_1 m_2 m_3} E_1 \Gamma^{m_1 m_2 m_3} \Gamma^9 \Gamma^0) \n
&& \qquad \qquad \qquad \qquad
\times (\Gamma^0 \tilde{D}_0 \Theta + iv\Gamma^9 [\Phi, \Theta]).
\end{eqnarray}
This algebra closes among the supertransformation, the translation and the gauge transformation on-shell. 

This dynamical supersymmetry and kinematical supersymmetry forms algebra generated by 32 supercharges, which is consistent with M-theory. 


%
%
%
%

\section{Conclusion and Discussion}
\setcounter{equation}{0}
In this paper, we have extended the BFSS matrix theory and obtained the 3-algebra BFSS matrix theory. The model allows any Lie 3-algebra. It owns the same supersymmetries as the original theory, thus as the IMF limit of M-theory.

As a first step to study dynamics of the extended model, we have chosen the minimal Lie 3-algebra that includes $u(N)$ and studied dynamics of the model. There are exactly two phases because of the constraint (\ref{constraint}). While the model reduces to the original BFSS matrix theory in one phase,  it reduces to the new simple supersymmetric action (\ref{BFSSlastaction}) in the other phase.

While we have obtained the extended model admitting any Lie 3-algebra, we have chosen the minimal Lie 3-algebra in this paper. In general, the extended model is not equivalent to the original matrix model. For example, the fifth algebra in (\ref{full3-alg}) gives a non-vanishing potential for $\Phi_{\alpha}$ and $X^M_{\alpha}$. The extended model should own a variety of new dynamics.

\vspace*{1cm}

\section*{Acknowledgements}
We would like to thank T. Asakawa, K. Hashimoto, N. Kamiya, H. Kunitomo, T. Matsuo, S. Moriyama, K. Murakami, J. Nishimura, S. Sasa, F. Sugino, T. Tada, S. Terashima, S. Watamura, K. Yoshida, and especially H. Kawai and A. Tsuchiya for valuable discussions.


\vspace*{0cm}




\begin{thebibliography}{99}



\bibitem{BFSS}T. Banks, W. Fischler, S.H. Shenker, L. Susskind, ``M Theory As A Matrix Model: A Conjecture,"Phys.Rev.{\bf D55}(1997) 5112, arXiv:hep-th/9610043.


\bibitem{BLG1}J. Bagger, N. Lambert, ``Modeling Multiple M2's," Phys.Rev.{\bf D75}:045020,2007, arXiv:hep-th/0611108.

\bibitem{Gustavsson}A. Gustavsson, ``Algebraic structures on parallel M2-branes," Nucl.Phys.{\bf B811}:66,2009, arXiv:0709.1260 [hep-th].

\bibitem{BLG2}J. Bagger, N. Lambert, ``Gauge Symmetry and Supersymmetry of Multiple M2-Branes," Phys.Rev.{\bf D77}:065008,2008, arXiv:0711.0955 [hep-th].

\bibitem{Lorentz0}S. Mukhi, C. Papageorgakis, ``M2 to D2," JHEP{\bf 0805}:085,2008, arXiv:0803.3218 [hep-th].

\bibitem{Iso}Y. Honma, S. Iso, Y. Sumitomo, S. Zhang, ``Janus field theories from multiple M2 branes," Phys.Rev.{\bf D78}:025027,2008, arXiv:0805.1895 [hep-th].
\bibitem{ABJM}O. Aharony, O. Bergman, D. L. Jafferis, J. Maldacena, ``N=6 superconformal Chern-Simons-matter theories, M2-branes and their gravity duals," JHEP{\bf 0810}: 091, 2008, arXiv: 0806.1218 [hep-th]

\bibitem{N=6BL}J. Bagger, N. Lambert, ``Three-Algebras and N=6 Chern-Simons Gauge Theories," Phys. Rev.{\bf D79}: 025002, 2009, arXiv: 0807.0163 [hep-th]



\bibitem{Filippov}V. T. Filippov, n-Lie algebras, Sib. Mat. Zh. 26, No. 6, (1985) 126140.

\bibitem{BergshoeffSezginTownsend}E. Bergshoeff, E. Sezgin, P.K. Townsend, Supermembranes and Eleven-Dimensional Supergravity, Phys. Lett.  B189 (1987) 75.


\bibitem{deWHN}B. de Wit, J. Hoppe, H. Nicolai, On the Quantum Mechanics of Supermembranes, Nucl. Phys.  B305 (1988) 545.

\bibitem{Nogo1}J. Figueroa-O'Farrill, G. Papadopoulos, ``Pluecker-type relations for orthogonal planes," J. Geom. Phys. 49 (2004) 294, math/0211170]


\bibitem{Jabbari}M. M. Sheikh-Jabbari, Tiny Graviton Matrix Theory: DLCQ of IIB Plane-Wave String Theory, A Conjecture , JHEP  0409 (2004) 017,  arXiv:hep-th/0406214.


\bibitem{GomisSalimPasserini}J. Gomis, A. J. Salim, F. Passerini, Matrix Theory of Type IIB Plane Wave from Membranes, JHEP  0808 (2008) 002,  arXiv:0804.2186.


\bibitem{HosomichiLeeLee}K. Hosomichi, K. Lee, S. Lee, Mass-Deformed Bagger-Lambert Theory and its BPS Objects, Phys.Rev.  D78 (2008) 066015,  arXiv:0804.2519.


\bibitem{Nogo2}G. Papadopoulos, ``M2-branes, 3-Lie Algebras and Plucker relations," JHEP 0805 (2008) 054, arXiv:0804.2662 [hep-th]



\bibitem{sp1}D. Gaiotto, E. Witten, Janus Configurations, Chern-Simons Couplings, And The Theta-Angle in N=4 Super Yang-Mills Theory, arXiv:0804.2907[hep-th].

\bibitem{sp2}K. Hosomichi, K-M. Lee, S. Lee, S. Lee, J. Park, N=5,6 Superconformal Chern-Simons Theories and M2-branes on Orbifolds, JHEP  0809 (2008) 002.



\bibitem{Nogo3}J. P. Gauntlett, J. B. Gutowski, ``Constraining Maximally Supersymmetric Membrane Actions," JHEP 0806 (2008) 053, arXiv:0804.3078 [hep-th] 

\bibitem{Lorentz1}J. Gomis, G. Milanesi, J. G. Russo, ``Bagger-Lambert Theory for General Lie Algebras," JHEP{\bf 0806}:075,2008, arXiv:0805.1012 [hep-th].

\bibitem{Lorentz2}S. Benvenuti, D. Rodriguez-Gomez, E. Tonni, H. Verlinde, ``N=8 superconformal gauge theories and M2 branes," JHEP{\bf 0901}:078,2009, arXiv:0805.1087 [hep-th].

\bibitem{Lorentz3}P.-M. Ho, Y. Imamura, Y. Matsuo, ``M2 to D2 revisited," JHEP{\bf 0807}:003,2008, arXiv:0805.1202 [hep-th].


\bibitem{sp3}M. Schnabl, Y. Tachikawa, Classification of N=6 superconformal theories of ABJM type, arXiv:0807.1102[hep-th].


\bibitem{PangWang}Y. Pang, T. Wang, From N M2's to N D2's, Phys. Rev.  D78 (2008) 125007,  arXiv:0807.1444.


\bibitem{ABJ}O. Aharony, O. Bergman, D. L. Jafferis, Fractional M2-branes, JHEP  0811 (2008) 043,  arXiv:0807.4924.


\bibitem{text}P. de Medeiros, J. Figueroa-O'Farrill, E. Me'ndez-Escobar, P. Ritter, ``On the Lie-algebraic origin of metric 3-algebras," Commun. Math. Phys. 290 (2009) 871, arXiv:0809.1086 [hep-th]

\bibitem{Nogo8}S. A. Cherkis, V. Dotsenko, C. Saeman, ``On Superspace Actions for Multiple M2-Branes, Metric 3-Algebras and their Classification," Phys. Rev. D79 (2009) 086002, arXiv:0812.3127 [hep-th]

\bibitem{NishinoRajpoot}H. Nishino, S. Rajpoot, Triality and Bagger-Lambert Theory, Phys. Lett.  B671 (2009) 415,  arXiv:0901.1173.

\bibitem{kac}P.-M. Ho, Y. Matsuo, S. Shiba, ``Lorentzian Lie (3-)algebra and toroidal compactification of M/string theory," arXiv: 0901.2003 [hep-th]

\bibitem{Class}P. de Medeiros, J. Figueroa-O'Farrill, E. Mendez-Escobar, P. Ritter, ``Metric 3-Lie algebras for unitary Bagger-Lambert theories," JHEP{\bf 0904}: 037, 2009, arXiv:  0902.4674 [hep-th] 

\bibitem{GustavssonRey}A. Gustavsson, S-J. Rey, Enhanced N=8 Supersymmetry of ABJM Theory on R(8) and R(8)/Z(2), arXiv:0906.3568 [hep-th].

\bibitem{HanadaMannelliMatsuo}M. Hanada, L. Mannelli, Y. Matsuo, Large-N reduced models of supersymmetric quiver, Chern-Simons gauge theories and ABJM, arXiv:0907.4937 [hep-th].

\bibitem{IshikiShimasakiTsuchiya}G. Ishiki, S. Shimasaki, A. Tsuchiya, Large N reduction for Chern-Simons theory on $S^3$, Phys. Rev.  D80 (2009) 086004,  arXiv:0908.1711.

\bibitem{KawaiShimasakiTsuchiya}H. Kawai, S. Shimasaki, A. Tsuchiya, Large N reduction on group manifolds, arXiv:0912.1456 [hep-th].

\bibitem{IshikiShimasakiTsuchiya2}G. Ishiki, S. Shimasaki, A. Tsuchiya, A Novel Large-N Reduction on $S^3$: Demonstration in Chern-Simons Theory, arXiv:1001.4917 [hep-th].


\bibitem{DeBellisSaemannSzabo}J. DeBellis, C. Saemann, R. J. Szabo, Quantized Nambu-Poisson Manifolds in a 3-Lie Algebra Reduced Model, JHEP  1104 (2011) 075,  arXiv:1012.2236.








\bibitem{Palmkvist}J. Palmkvist, ``Unifying N = 5 and N = 6," JHEP 1105 (2011) 088, arXiv:1103.4860.





\bibitem{Nambu}Y. Nambu, ``Generalized Hamiltonian dynamics," Phys.Rev.{\bf D7}:2405-2414,1973.

\bibitem{Yoneya}H. Awata, M. Li, D. Minic, T. Yoneya, ``On the Quantization of Nambu Brackets," JHEP{\bf 0102}(2001) 013, arXiv:hep-th/9906248.

\bibitem{Minic}D. Minic, ``M-theory and Deformation Quantization," arXiv:hep-th/9909022.

\bibitem{bosonicM}M. Sato, `` Covariant Formulation of M-Theory," Int. J. Mod. Phys. A24 (2009), 5019, arXiv:0902.1333 [hep-th]




\bibitem{MModel}M. Sato, Model of M-theory with Eleven Matrices, JHEP 1007 (2010) 026.

\bibitem{LorentzianM}M. Sato, Supersymmetry and the Discrete Light-Cone Quantization Limit of the Lie 3-algebra Model of M-theory, Phys. Rev.  D85 (2012), 046003.

\bibitem{ZariskiM}M. Sato, ``Zariski Quantization as Second Quantization,'' Phys. Rev. D85 (2012) 126012, arXiv:1202.1466 [hep-th].




\bibitem{Smolin1}L. Smolin, ``M theory as a matrix extension of Chern-Simons theory," Nucl.Phys.{\bf B591}(2000) 227, hep-th/0002009.
 
\bibitem{Smolin2}L. Smolin, ``The cubic matrix model and a duality between strings and loops," hep-th/0006137.

\bibitem{Azuma}T. Azuma, S. Iso, H. Kawai, Y. Ohwashi, ``Supermatrix Models," Nucl.Phys.{\bf B610}(2001) 251, hep-th/0102168.



\bibitem{Ghost-Free}M. A. Bandres, A. E. Lipstein, J. H. Schwarz, ``Ghost-Free Superconformal Action for Multiple M2-Branes," JHEP{\bf 0807}: 117, 2008, arXiv:
0806.0054 [hep-th].




\bibitem{MSUSY}T. Banks, N. Seiberg, S. Shenker, ``Branes from Matrices," Nucl. Phys. {\bf B490}: 91, 1997, arXiv: hep-th/9612157.

















\end{thebibliography}
\end{document}